\documentclass[prd,aps,showpacs,nofootinbib]{revtex4}
\usepackage{amssymb}
\usepackage{graphicx}
\usepackage{amsmath}

\newcommand{\be}{\begin{equation}}
\newcommand{\ee}{\end{equation}}
\newcommand{\bq}{\begin{eqnarray}}
\newcommand{\eq}{\end{eqnarray}}

\begin{document}

\title{N=1-Supersymmetric Description of a Spin-1/2 charged Particle in a
5d-World}
\author{H. Belich $^{a,e}$, D. Cocuroci$^{b}$, G. S. Dias$^{c,d}$, J.A. Helay%
\"{e}l-Neto$^{b,e}$, M.T.D. Orlando$^{a,e}$ }
\affiliation{$^{a}${\small {Universidade Federal do Esp\'{\i}rito Santo (UFES),
Departamento de F\'{\i}sica e Qu\'{\i}mica, Av. Fernando Ferrari 514, Vit%
\'{o}ria, ES, CEP 29060-900, Brasil}}}
\affiliation{$^{b}${\small {CBPF - Centro Brasileiro de Pesquisas F\'{\i}sicas, Rua
Xavier Sigaud 150, Rio de Janeiro, RJ, CEP 22290-180, Brasil}}}
\affiliation{{\small {~}}$^{c}${\small Instituto Federal do Esp\'{\i}rito Santo (IFES) -
Campus Vit\'{o}ria,Av. Vit\'{o}ria 1729, Jucutuquara, Vit\'{o}ria - ES,
29040-780, CEP 29040-780, Brasil}}
\affiliation{{\small {~}}$^{d}${\small {Theoretical Physics Institute, Department of
Physics, University of Alberta, Edmonton, Alberta, Canada T6G 2J1}}}
\affiliation{{\small {~}}$^{e}${\small {Grupo de F\'{\i}sica Te\'{o}rica Jos\'{e} Leite
Lopes, C.P. 91933, CEP 25685-970, Petr\'{o}polis, RJ, Brasil}}}
\email{belichjr@gmail.com, denis.cocuroci@gmail.com, gilmar@ifes.edu.br,
helayel@cbpf.br, orlando@cce.ufes.br}

\begin{abstract}
We study the dynamics of a charged spin-$\frac{1}{2}$ particle in an
external 5-dimensional electromagnetic field. We then consider that we are
at the $TeV$\ scale, so that we can access the fifth dimension and carry out
our physical considerations in a $5-$dimensional brane. In this brane, we
focus our attention to the quantum-mechanical dynamics of a charged particle
minimally coupled to the $5-$dimensional electromagnetic field. We propose a
way to\ identify the Abraham-Lorentz back reaction force as an effect of the
extra ( fifth ) dimension. Also, a sort of dark matter behavior can be
identified in a particular regime of the dynamics of the particle
interacting with the bulk electric field.
\end{abstract}

\maketitle

Supersymmetry (SUSY) has emerged as a viable response to obtain a unified
picture of all interactions of Nature. The strengths of the three forces of
the Standard Model evolve to similar values at high temperatures, but never
converge to the same magnitude of the interactions.\ Including SUSY, the
strengths of the three forces of the Standard Model (SM)\textrm{\ }converge
to very close values at energies of the order of $10^{16}GeV$. With the
superpartner particles predicted by SUSY, the three forces approach the same
strengths at very high temperatures, making grand-unification (GUT)
possible.\ This extension is known as the Minimal Supersymmetric Standard
Model (MSSM).

Another aspect that SUSY contributes is a way-out to the gauge hierarchy
problem, i.e., it justifies why the energy of electroweak symmetry breaking
is so low as compared to the grand-unification scale. SUSY controls very
well the divergences which emerge in the SM and take the breaking scale to
the GUT region .

Due to discoveries from Wilkinson Microwave Anisotropy Probe (WMAP), among
others experiments, we have to include additional matter to the models of
the universe - matter, whose nature we do not know and used to denominate as
dark matter. Supersymmetry provides a natural explanation to dark matter if
we take hand of SUSY breaking mechanism. We have a splitting of mass and
this dark matter is not observed in current accessible energy scales.\ \ If
Supersymmetry is, in fact, a symmetry of Nature, what we see today must be
its low-energy remnant.\textsl{\ }Thus we can attribute the absence of
supersymmetry through the breaking accomplished by means of \ mechanisms 
\cite{susyvinculosquebra}, \cite{spinsusyB}, \cite%
{susyquebramecanicaquantica}, such that, a field theory should be related to
a supersymmetric Schr\"{o}dinger equation in each particle number sector of
the theory.

To study\textrm{\ }these aspects of Supersymmetric Quantum Field Theory at
low energies (non-relativistic limit) became a topic of special interest 
\cite{monopolosusyspinB}, \cite{aplicacaosusy}, \cite{capitulolivrosusy}. In
fact as we are interested in the regime in which particles are not generated
or destroyed, therefore we deal with non relativistic limit, with
Supersymmetric Quantum Mechanics description.

On the other hand, the introduction of extra dimensions to explain the
fundamental forces of Nature remains to the beginnings of General
Relativity. The Kaluza-Klein Theory try to explain the Gravity from
Electromagnetism in a space-time with more than four dimensions \cite{kaluza}%
.\textsl{\ }Nowadays, multidimensional space-time approach have appeared
connected with string models \cite{string}, and branes \cite{ruth, randal},\
as a possible mechanism to overcome the hierarchy problem. This way of
dealing with the problem is based on the idea that matter is restricted to a
\ $3-D$ brane immersed in a manifold with more spatial dimensions. In this
context, the extra dimensions are not compactified; this opens up the
possibility that we can have experimental detections \ \cite{exp} observing
directly the particle production at $TeV$ scale.

To investigate the possibility that QED\ in $(1+3)-D$, embedded in a
submanifold of a $(1+4)-D$\ brane, in this letter, we study Maxwell%
\'{}%
s electrodynamics in the $5-D$\ scenario. The analysis of this dynamics in $%
(1+4)-D$\ dimensions gives us the possibility to describe a hidden sector of
Electromagnetism which we can conjecture as dark energy \cite{dark, dark1,
dark2, dark3}, and a particle coupled to this field is a candidate to dark
matter.

One question we address to in our study of the motion of a charged particle
in a 5-dimensional supersymmetric scenario consists in checking how the
electromagnetic interactions can control its motion in the \ 3-dimensional
spatial sector corresponding to our world or may take it to the bulk \cite%
{randal}. If it is driven to the bulk , $(x^{1}=x^{2}=x^{3}=0,x^{4}\neq 0)$,
we can understand under what circumstances the external electromagnetic
fields in $5-D$ determine that the particle behaves as dark matter. On the
other hand, this discussion may guide us to relate the back radiation force
in $(1+3)-D$ with the mechanical power of the magnetic force in $5-D$.

In this letter, we are going to pursue an investigation of a Maxwell
electrodynamics in $(1+4)$ dimensions and, in this frame, we shall be
discussing particular aspects of the dynamics of charged massive particles
under the action of the $5-$dimensional electric and magnetid fields. We
pick up a particle of mass $m$ and spin-$1/2$ in an external field, \cite%
{monopolospinsusyA}, \cite{monopolosusyC}, \cite{ideiainicial},\cite%
{spinemsusy}. We build up a superspace action for this model,\textrm{\ }read
off the supersymmetry.transformations of the component coordinates. At the
end, we study the dynamics of a particle in a $(1+4)-$dimensional brane.

The scenario we draw to pursue our investigation is as follows: we adopt the
viewpoint of extra dimensions as large as $(TeV)^{-1}$, according to the
frame established by Dvali et al. \cite{dark2}. We then consider that we are
at the $TeV$\ scale, so that we can access the fifth dimension and carry out
our physical considerations in a $5-$dimensional brane. In this brane, we
focus our attention to the quantum-mechanical dynamics of a charged particle
minimally coupled to the $5-$dimensional electromagnetic field. Clearly,
this particle must be massive enough to an extent that it still makes sense
to consider its dynamics described by Quantum Mechanics. Then, we suppose
that the mass, $m$, of our (charged) test particle is of the $TeV-$order,
but still larger than the energy scale we are considering. This means that
there is not enough energy to penetrate the Compton wavelenght of the
particle, so that a quantum-mechanical approach makes sense. So, we are
considering a picture such that a charged massive particle is studied at an
energy scale $\varepsilon $\ ( $\varepsilon \leq 2m$) for which the fifth
dimension shows up and, then, the dynamics of the particle is governed by a
quantum-mechanical treatment and the particle feels the action of the $5-$%
dimensional Maxwell field, which encompasses the electric $(\vec{E})$\ and
magnetic $(\vec{B})$\ fields of the ordinary $4-$dimensional Maxwell theory,
and includes two extra electric- and magnetic-like fields genuinely
connected to the fifth dimension.

Supersymmetry (actually Supersymmetric Quantum Mechanics) is present in our
approach for we know that the treatment of spin$-\frac{1}{2}$\ particles may
be associated to a supersymmetric approach in which the spin variables are
identified with the Grassmannian partner of the variables describing the
particle position. The Supersymmetry we take about here is not a high-energy
SUSY; it is simply the sort of dynamical symmetry which is underneath the
description of the quantum-mechanics aspects of $s=\frac{1}{2}$-particles.
We quote relevant references for such a discussion in \cite{cooper}.

Once our physical framework has been settled down, we start our
considerations on the dynamics of our test particle under the action of the
5-dimensional electromagnetic field.

We intend to investigate the $5-D$ electrodynamics. The Maxwell%
\'{}%
s electromagnetism in 4-dimensional space-time could be the derivative of a
more fundamental theory in $5-D$ dimensions. This electrodynamics lives in a 
$5-D$ hypersurface of a brane. Vectors and tensors shall be decomposed in
terms of $SO(3)$ indexes; we split the $4$-th component which behave as a
scalar under $SO(3)$ rotations.

We consider the following action in $1+4$ dimensions,

\begin{equation}
L=\overline{\Psi }(i\Gamma ^{\mu }D_{\mu }-m)\Psi ;  \label{Ldirac5d}
\end{equation}

where we define: 
\begin{eqnarray}
\mu &\in &\{0,1,2,3,4\},\quad \eta =(+,-,-,-,-),\quad  \label{Definicao1} \\
x^{\mu } &=&(t,x=x^{1},y=x^{2},z=x^{3},x^{4}),\quad F_{\mu \nu }=\partial
_{\mu }A_{\nu }-\partial _{\nu }A_{\mu } \\
D_{_{\mu }} &=&\partial _{\mu }+ieA_{\mu },\quad A_{\mu
}=(A_{0},A_{i},A_{4}),\quad \left\{ \Gamma ^{\mu },\Gamma ^{\nu }\right\}
=2\eta ^{\mu \nu }, \\
\partial ^{i} &\Longleftrightarrow &-\overrightarrow{\nabla },\quad \partial
_{0}\Longleftrightarrow \frac{\partial }{\partial t},\quad
E^{i}\Longleftrightarrow \overrightarrow{E},\quad B^{i}\Longleftrightarrow 
\overrightarrow{B},\quad , \\
F_{04} &=&-\mathcal{E},\quad F_{i4}=\mathcal{B}_{i},\quad F_{ij}=-\epsilon
_{ijk}B_{k},\quad F_{0i}=E_{i},
\end{eqnarray}

where $i,j,k\in \{1,2,3\}$. Notice that the scalar, $\mathcal{E}$, and the
vector, $\mathcal{\vec{B}}$, accompany the vectors $\overrightarrow{E}$ and $%
\overrightarrow{B}$, later on to be identified with the electric and
magnetic fields, respectively. Our explicit representation for the
gamma-matrices is given in the Appendix. We set: 
\begin{equation}
\Psi =\left( 
\begin{array}{c}
\psi _{1} \\ 
\psi _{2} \\ 
\psi _{3} \\ 
\psi _{4}%
\end{array}%
\right) ,\quad \overline{\Psi }=\Psi ^{\dagger }\Gamma ^{0}
\end{equation}

We now take the equation of motion for ${\Psi }$ from the Euler-Lagrange
equations (\ref{Ldirac5d}). We suppose a stationary solution, 
\begin{equation}
\Psi (x,t)=\exp (-i\epsilon t)\chi (x);
\end{equation}

here, we are considering that the external field does not depend on t, and $%
A_{0}=0$.

To observe the properties of the matter at low energy scale we have to
obtain the non-relativistic limits. Then, considering the non-relativistic
limit of the reduced theory, the equations of motions read:

\begin{eqnarray}
(\epsilon -m)\chi _{1}+(i\sigma ^{i}(\partial _{i}+ieA_{i})+\left( \partial
_{4}+ieA_{4}\right) )\chi _{2} &=&0,  \label{eqespinorial} \\
(i\sigma ^{i}(\partial _{i}+ieA_{i})-\left( \partial _{4}+ieA_{4}\right)
)\chi _{1}+(\epsilon +m))\chi _{2} &=&0.
\end{eqnarray}%
therefore, we see that $\chi =\left( 
\begin{array}{c}
\chi _{1} \\ 
\chi _{2}%
\end{array}%
\right) $ looses two degrees of freedom, described by the "weak" spinor, $%
\chi _{2}$. So the Pauli-like Hamiltonian we get to reads as below:

\begin{equation}
H=-\frac{1}{2m}\left[ \left( \overrightarrow{\nabla }-ie\overrightarrow{A}%
\right) ^{2}+\left( \partial _{4}-ieA^{4}\right) ^{2}+e\overrightarrow{%
\sigma }\left( \overrightarrow{\nabla }\times \overrightarrow{A}+2\mathcal{%
\vec{B}}\right) +e\phi \right]  \label{Hpauli54dn1}
\end{equation}

where we used (\ref{Definicao1}).

If we define, $p_{i}=i\overrightarrow{\nabla }$ ($\hbar =1$), then the
Hamiltonian (\ref{Hpauli54dn1}) will become

\begin{eqnarray}
H &=&\frac{1}{2m}\left[ \left( \vec{p}-e\vec{A}\right) ^{2}+\left(
p^{4}-eA^{4}\right) ^{2}\right] -e\left( \vec{B}+2\mathcal{\vec{B}}\right) 
\vec{S}+e\phi ; \\
\vec{S} &=&\frac{e}{2m}\vec{\sigma},
\end{eqnarray}

and we obtain the Pauli Hamiltonian revealing the extra-dimension
contribution.

Now it could be suitable to work out some specific aspect of Classical
Electrodynamics in $(1+4)-D$.

We shall present the whole set of fields as written in terms of $SO(3)-$%
tensor representations and to put in a manifest form Maxwell equations in
five dimensions, but written in terms of the $SO(3)$ fields. Next, we
written down the components of the energy-momentum tensor present its
associated continuity equations and the meaning of the $\Theta _{\mu \nu }-$%
components. This tasks may be clarifying for the sake of our final
discussion on the study of the classical dynamics of a $5-D$ charged
particle in an electromagnetic field.

We start by considering the equations of motion $\epsilon ^{\mu \rho \sigma
\beta \gamma }\partial _{\sigma }F_{\beta \gamma }=0$ and $\partial _{\sigma
}F^{\sigma \gamma }=\rho ^{\gamma }$; where $F_{\beta \gamma }$ is the field
strength defined in (\ref{Definicao1}) and $\epsilon ^{\mu \rho \sigma \beta
\gamma }$ is the Levi-Civita tensor in 5 dimensions. Then, we have:

\begin{eqnarray}
\overrightarrow{\nabla }\cdot \overrightarrow{E}-\frac{\partial \mathcal{E}}{%
\partial x^{4}} &=&0, \\
\overrightarrow{\nabla }\times \overrightarrow{B}-\frac{\partial 
\overrightarrow{E}}{\partial t}-\frac{\partial \mathcal{\vec{B}}}{\partial
x^{4}} &=&0, \\
\frac{\partial \mathcal{E}}{\partial t}+\overrightarrow{\nabla }\cdot 
\mathcal{\vec{B}} &=&0, \\
\overrightarrow{\nabla }\times \overrightarrow{E} &=&-\frac{\partial 
\overrightarrow{B}}{\partial t}, \\
\frac{\partial \mathcal{\vec{B}}}{\partial t}\overrightarrow{+\nabla }\cdot 
\mathcal{E}+\frac{\partial \overrightarrow{E}}{\partial x^{4}} &=&0, \\
\overrightarrow{\nabla }\times \mathcal{\vec{B}}-\frac{\partial 
\overrightarrow{B}}{\partial x^{4}} &=&0, \\
\overrightarrow{\nabla }\cdot \overrightarrow{B} &=&0.
\end{eqnarray}

We observe that,

\begin{eqnarray}
\overrightarrow{E} &\Longleftrightarrow &F_{0i}\equiv \partial
_{0}A_{i}-\partial _{i}A_{0}=0;  \label{definicao2} \\
F_{04} &\equiv &\partial _{0}A_{4}-\partial _{4}A_{0}=-\mathcal{E}; \\
F_{ij} &=&\partial _{i}A_{j}-\partial _{j}A_{i}\equiv -\epsilon _{ijk}B_{k};
\\
F_{i4} &=&\partial _{i}A_{4}-\partial _{4}A_{i}\equiv \mathcal{B}_{i}.
\end{eqnarray}

with these equations we can better understand the influence of the fifth
dimension in the particle motion. From the energy-momentum tensor, $\Theta
^{\mu }\,_{\kappa }=F^{\mu \alpha }F_{\alpha \kappa }+\frac{1}{4}\delta
_{\kappa }^{\mu }F^{\alpha \beta }F_{\alpha \beta }$, evaluating each set of
separate components, we have::

\begin{equation}
\begin{array}{ccc}
\Theta ^{0}\,_{0} & = & \frac{1}{2}[(E^{2})+B^{2}+(\mathcal{E}^{2})+\mathcal{%
B}^{2}],%
\end{array}%
\end{equation}

\subsubsection*{%
\protect\begin{equation}
\protect\begin{array}{ccl}
\Theta ^{0}\,_{i} & = & -[(\vec{E}\times \vec{B})+\mathcal{E\vec{B}}]_{i},%
\protect\end{array}%
\protect\end{equation}%
}

\begin{equation}
\begin{array}{ccl}
\Theta ^{i}\,_{j} & = & E_{i}E_{j}+B_{i}B_{j}-e_{i}e_{j}-\frac{1}{2}\delta
_{j}^{i}(E^{2}+B^{2}+\mathcal{E}^{2}-\mathcal{B}^{2}),%
\end{array}%
\end{equation}

\begin{equation}
\begin{array}{ccl}
\Theta ^{0}\,_{4} & = & \vec{E}\cdot \mathcal{\vec{B}},%
\end{array}%
\end{equation}

\begin{equation}
\begin{array}{ccl}
\Theta ^{i}\,_{4} & = & -[\mathcal{E}\vec{E}+\mathcal{\vec{B}}\times \vec{B}%
]_{i},%
\end{array}%
\end{equation}%
{}

\begin{equation}
\begin{array}{ccc}
\Theta ^{4}\,_{4} & = & -\frac{1}{2}[(E^{2})-B^{2}+\mathcal{B}^{2}-(\mathcal{%
E}^{2})]%
\end{array}%
\end{equation}%
which is interpreted as a new "density" of pressure toward our extra
dimension $s$. In our study the 5-D Electromagnetics this Poynting's theorem
is summarized by the following expression:

\begin{equation}
\frac{\partial }{\partial t}u+\nabla \cdot \vec{S}+\frac{\partial }{\partial
s}\xi =-\vec{j}\cdot \vec{E}+j_{s}\mathcal{E},
\label{eq:teorema de poynting}
\end{equation}

where $S$ is the "Poynting vector" representing the energy flow, $%
\overrightarrow{j}$ is the current density and $\overrightarrow{E}$ is the
electric field, and $u$ is the density of electromagnetic energy.

\begin{equation*}
\begin{array}{ccl}
u & = & \frac{1}{2}[(E^{2})+B^{2}+(\mathcal{E}^{2})+\mathcal{B}^{2}], \\ 
\vec{S} & = & (\vec{E}\times \vec{B})+\mathcal{E\vec{B}}, \\ 
\xi & = & -\overrightarrow{E}\cdot \mathcal{\vec{B}}.%
\end{array}%
\end{equation*}

The $5-D$ expression for the Lorentz force follows in connection with the
following \ continuity equation ( in presence of external source ):

\begin{equation}
\frac{1}{c}\frac{\partial \overrightarrow{S}}{\partial t}-\nabla \cdot
\sigma +\frac{\partial \overrightarrow{\chi }}{\partial s}=-\rho 
\overrightarrow{E}-\overrightarrow{j}\times \overrightarrow{B}+j_{s}\mathcal{%
\vec{B}}
\end{equation}

and

\begin{equation}
\begin{array}{ccl}
\vec{S} & = & (\vec{E}\times \vec{B})+\mathcal{E\vec{B}}, \\ 
\sigma & = & \Theta ^{i}\,_{j}, \\ 
\vec{\chi} & = & (\frac{1}{c^{2}}\mathcal{E}\vec{E}+\mathcal{\vec{B}}\times 
\vec{B}).%
\end{array}%
\end{equation}

In our scenario, 5-D, there is a third term which embraces the conservation
of scalar moments,

\begin{equation}
-\frac{1}{c^{2}}\frac{\partial \xi }{\partial t}-\nabla \cdot \vec{\chi}+%
\frac{\partial }{\partial s}\Omega =-\rho b+\vec{j}\cdot \mathcal{\vec{B}}
\label{eq: new force}
\end{equation}

and

\begin{equation}
\begin{array}{ccl}
\xi & = & -\vec{E}\cdot \mathcal{\vec{B}}, \\ 
\Omega & = & -\frac{1}{2}[(E^{2}-B^{2}-(\mathcal{E}^{2})+\mathcal{B}^{2}].%
\end{array}%
\end{equation}

Now, we are back to the situation of a charged particle under the action of
an external $5-D$ eletromagnetic field. We shall take this particular system
in order to better understand how the $5-D$ fields may act upon charged
particles dynamics may take place partly in $(1+3)-D$. This will open up for
us some interesting discussions on the back reaction of force and may even
lead us to an interpretation of a possible sort of dark-matter-like charged
particles.

To render more systematic to our discussion, we think it is advisable to set
up a superfield approach. We can define the N=1-supersymmetric model in
analogy with the model presented above, eq.(\ref{Lpauli54dn1spin}). It is
not a trivial task, and to solve this question we start by defining the
superfields as below:

\begin{equation}
\Phi _{i}(t,\theta )=x_{i}(t)+i\theta \psi _{i}(t),\Phi _{4}(t,\theta
)=x_{4}(t)+i\theta \psi _{4}(t),  \label{Definicao3}
\end{equation}

The supercharge operators and the covariant derivatives are given by: 
\begin{equation}
Q=\partial _{\theta }+i\theta \partial _{t}\quad D=\partial _{\theta
}-i\theta \partial _{t}\quad \text{and }H=i\partial _{t}.
\label{Operadoresn1}
\end{equation}

We have to set up a Lagrangian in terms of the $\Phi _{i}$'s and $\Phi _{4}$
so as to recover the Hamiltonian of eq. (\ref{Hpauli54dn1}). The
N=1-supersymmetric Lagrangian that generates the appropriate bosonic sector
is given by: 
\begin{equation}
\mathcal{L}_{1}=\frac{i}{2}m\left( \dot{\Phi}_{i}D\Phi _{i}+\dot{\Phi}%
_{4}D\Phi _{4}\right) +ie\left[ (D\Phi _{i})A_{i}(\Phi _{j},\Phi
_{4})+\left( D\Phi _{4}\right) A_{4}(\Phi _{j},\Phi _{4})\right] .
\label{Lsusyn1}
\end{equation}%
Integrating over the Grassman variable, we obtain:

\begin{equation}
L_{1}=\frac{1}{2}m\left( (\dot{x}_{i})^{2}+(\dot{x}_{4})^{2}\right) -\frac{i%
}{2}\left( \psi _{i}\dot{\psi}_{i}+\psi _{4}\dot{\psi}_{4}\right) +e\left(
A_{i}\dot{x}_{i}+A_{4}\dot{x}_{4}\right) +e\phi -\frac{ie}{2}\left( B_{i}+2%
\mathcal{B}_{i}\right) \epsilon _{ijk}\psi _{j}\psi _{k},
\label{Lpauli54dn1}
\end{equation}%
where the dot stands for a derivative with respect to time. We can also write

\begin{equation}
L_{1}=\frac{1}{2}m\left( (\dot{x}_{i})^{2}+(\dot{x}_{4})^{2}\right) -\frac{i%
}{2}\left( \psi _{i}\dot{\psi}_{i}+\psi _{4}\dot{\psi}_{4}\right) +e\left(
A_{i}\dot{x}_{i}+A_{4}\dot{x}_{4}\right) +e\phi +eB_{i}S_{i}+e2\mathcal{B}%
_{i}S_{i},  \label{Lpauli54dn1spin}
\end{equation}%
where we define the spin by the product below:

\begin{equation}
S_{i}=-\frac{i}{2}\epsilon _{ijk}\psi _{j}\psi _{k}.
\end{equation}

\begin{eqnarray}
H_{1} &=&\frac{1}{2}m\left( (\dot{x}_{i})^{2}+(\dot{x}_{4})^{2}\right)
+i\left( \psi _{i}\dot{\psi}_{i}+\psi _{4}\dot{\psi}_{4}\right) +e\phi +%
\frac{ie}{2}\left( B_{i}+2\mathcal{B}_{i}\right) \epsilon _{ijk}\psi
_{j}\psi _{k}, \\
H_{1} &=&\frac{1}{2m}\left[ \left( \vec{p}-e\vec{A}\right) ^{2}+\left(
p^{4}-eA^{4}\right) ^{2}\right] +\frac{ie}{2}\left( B_{i}+2\mathcal{B}%
_{i}field\right) \epsilon _{ijk}\psi _{j}\psi _{k}+e\phi +i\left( \psi _{i}%
\dot{\psi}_{i}+\psi _{4}\dot{\psi}_{4}\right) ,
\end{eqnarray}

where we observe a new Pauli coupling with the field $\mathcal{B}_{i}$ and
the contributions of the fermionic coordinates $\psi _{j}$ and $\psi _{4}$.

The eqs. of motion in the fermionic sector are:

\begin{eqnarray}
\dot{\psi}_{4} &=&0,\text{ } \\
\dot{\psi}_{i} &=&e\left( \vec{B}+2\mathcal{\vec{B}}\right) _{j}\epsilon
_{ijk}\psi _{\kappa },
\end{eqnarray}

where it is manifest the Pauli-type coupling in the fermionic sector. For
the coordinates $\vec{x}$ and $x^{4}$, the equations of motion assume the
form:

\begin{eqnarray}
m\frac{\partial ^{2}\overrightarrow{x}}{\partial t^{2}} &=&e\overrightarrow{v%
}\times \overrightarrow{B}-e\mathcal{\vec{B}}\dot{x}^{4}+e\vec{E},
\label{a1} \\
m\frac{\partial ^{2}x^{4}}{\partial t^{2}} &=&-e\mathcal{\vec{B}}.%
\overrightarrow{v}-e\mathcal{E},  \label{a2}
\end{eqnarray}

where the extended Lorentz force gets contribution from $\mathcal{\vec{B}}$.
To focus the possible novelties that this model may reveal, we have to pay
attention to the extra-dimension contribution. To understand its exclusive
effect, we set $\vec{E}=0$, $\overrightarrow{B}=0$, and $\mathcal{E}=0;$
then, we get:

\begin{eqnarray}
m\frac{\partial ^{2}\overrightarrow{x}}{\partial t^{2}} &=&-e\mathcal{\vec{B}%
}\dot{x}_{4},  \label{t1} \\
m\frac{\partial ^{2}x^{4}}{\partial t^{2}} &=&-e\mathcal{\vec{B}}.%
\overrightarrow{v}.  \label{t2}
\end{eqnarray}

Manipulating the equations above, we obtain:

\begin{eqnarray}
\frac{\partial ^{3}x^{4}}{\partial t^{3}} &=&\alpha ^{2}.\dot{x}_{4}, \\
\alpha &=&\frac{e}{m}\left\vert \mathcal{\vec{B}}\right\vert .
\end{eqnarray}

The solutions are:

\begin{eqnarray*}
(i)\text{ \ \ }\dot{x}^{4} &\sim &\exp \left( +\alpha t\right) , \\
(ii)\text{ \ }\dot{x}^{4} &\sim &\exp \left( -\alpha t\right) .
\end{eqnarray*}

Taking into account the run-away solution $(i),$ and replacing in (\ref{t1}%
), we obtain:

\begin{eqnarray}
m\frac{\partial ^{2}\overrightarrow{x}}{\partial t^{2}} &\sim &-\mathcal{%
\vec{B}}\exp \left( +\alpha t\right) , \\
\overrightarrow{x} &\sim &-\mathcal{\vec{B}}\exp \left( +\alpha t\right) .
\end{eqnarray}

In more explicit way, we have:

\begin{equation*}
F_{+}=m\frac{\partial ^{2}\overrightarrow{x_{+}}}{\partial t^{2}}=-e\mathcal{%
\vec{B}}\left\vert \alpha \right\vert \exp \left( +e\frac{\left\vert 
\mathcal{\vec{B}}\right\vert }{m}t\right) ,
\end{equation*}

In this case, the particle moves in the opposite direction of $\mathcal{\vec{%
B}}$ and has an exponential growth force $F_{+}$

\begin{equation*}
F_{-}=m\frac{\partial ^{2}\overrightarrow{x_{-}}}{\partial t^{2}}=e\mathcal{%
\vec{B}}\left\vert \alpha \right\vert \exp \left( -e\frac{\left\vert 
\mathcal{\vec{B}}\right\vert }{m}t\right) ,
\end{equation*}

In this case, the particle moves in the same direction of $\mathcal{\vec{B}}$
and has a decreasing exponential strength $F_{-}$.

By (\ref{t1}), we can suppose that the extra term in the Lorentz force $(%
\mathcal{\vec{B}}\dot{x}^{4})$ can be associated with,the back reaction or
the Abraham-Lorentz force. Then, we could interpret the back reaction force
as an effect of a fifth dimension of space-time where the $\mathcal{\vec{B}}$%
-field is the 3-dimensional projection of the true magnetic field in (1+4)-D$%
.$

On the other hand taking into account $(ii),$ and replacing in (\ref{t1}),
we obtain:

\begin{eqnarray}
m\frac{\partial ^{2}\overrightarrow{x}}{\partial t^{2}} &\sim &\mathcal{\vec{%
B}}\exp \left( -\alpha t\right) , \\
\overrightarrow{x} &\sim &\mathcal{\vec{B}}\exp \left( -\alpha t\right) .
\end{eqnarray}

But, if we study the case in which $\mathcal{\vec{B}}=0$, and $\mathcal{E}%
\neq 0,$\ from (\ref{a2}), we get:

\begin{equation*}
\frac{\partial ^{2}x^{4}}{\partial t^{2}}=-\frac{e\mathcal{E}}{m},
\end{equation*}

which shows that the particle indefinitely scapes to the extra dimension $%
x^{4}$. Here, we can conclude that the particle may describe some component
of dark matter, and the field $\mathcal{E}$, the piece remnant of the 5-D
electric field drives the particle away from the Minkowski brane.

So, to conclude, we would like to stress \ and comment on a few issues. We
have studied an $N=1,$ $D=5$-supersymmetric particle, in an external
electromagnetic field by considering the non-relativistic regime, which, as
already motivated in the introductory comments, is reasonable for we
consider heavy particles with low kinetic energy\textsl{.} We admit that
this model could be interesting to describe cold dark matter. We suppose
that this particle is immersed in a (1+4)-dimensional brane. In this
extended electromagnetic field, the ($\mathcal{B}_{i}$ and $\mathcal{E)}$%
-fields appear in the Maxwell equations. The effect of the fifth coordinate $%
\left( x^{4}\right) $ is felt by means of an extension of the Lorentz force
in $4$ dimensions. And, by focusing on the evolution \ of the $x^{4}$%
-coordinate, we have identified two possible relevant situations in
connection with the \ time evolution of the $x^{4}$-coordinate. In the
situation, $x^{4}$ is of the run-away type and we associate its effect as
the Abraham-Lorentz back reaction force in our $4-$dimensional world. By
adopting this point of view, we propose that the effect on an extra
dimension may show up under the guise of the back reaction force in the
dynamics of a charged particle subjected to an electromagnetic field.

On the other hand, the presence of an electric field may drive the charged
particle to the bulk and, by virtue of this mechanism, we propose that the
charged particle, in this regime, has a similar behaviour to the dark matter
particles. Of course, the particle is not electrically neutral; so, in this
sense, it cannot be a genuine dark matter constituent. However, the extra
electric field, also confined to the bulk, drives the test particle from the
Minkowski brane and, once in the bulk, it interacts with the bulk field $%
\mathcal{\vec{B}}$. Though charged, it escapes to the bulk and this is the
reason we say it similar to the dark matter particles. In connection with
this mechanism, we point out the results of the work of Ref.\cite{dimo},
where the authors discuss the stringent constraints on the existence of
charged dark matters.

The authors express their gratitude to CNPq for the financial support.


\begin{thebibliography}{99}
\bibitem{susyvinculosquebra} Edward Witten, Nucl. Phys. B 202 (1982) 253.

\bibitem{spinsusyB} P. Salomonson, J. W. Van Holten, Nucl. Phys. B 196
(1982) 509.

\bibitem{susyquebramecanicaquantica} D. Lancaster, IL Nuoevo Cimento 79 A, 1
(1984) 28.

\bibitem{monopolosusyspinB} Donald Spector, Phys.Lett. B474 (2000) 331.

\bibitem{aplicacaosusy} Lene Vestergaard Hau, J. A. Golovchenko, Michael M.
Burns, Phys. Rev. Lett. 74 16 (1995) 3138.

\bibitem{capitulolivrosusy} Ashok Das, Field Theory a path integral
approach, World Scientific Publishing - Vol 52 - Chapter 6.

\bibitem{kaluza} Abdus Salam and J. A. Strathdee. On Kaluza-Klein Theory.
Annals Phys., 141:316--352,1982.

\bibitem{string} Jan de Boer. String theory: An update. Nucl. Phys. Proc.
Suppl., 117:353--372, 2003.

\bibitem{ruth} Ruth Durrer. Braneworlds. AIP Conf. Proc., 782:202--240, 2005.

\bibitem{randal} Lisa Randall and Raman Sundrum. A large mass hierarchy from
a small extra dimension,.Phys. Rev. Lett., 83:3370--3373, 1999.

\bibitem{exp} V. A. Rubakov. Large and infinite extra dimensions: An
introduction. Phys. Usp., 44:871--893, 2001.

\bibitem{dark} S. Perlmutter et al. Measurements of Omega and Lambda from 42
High-Redshift Supernovae.Astrophys. J., 517:565--586, 1999.

\bibitem{dark1} G. Bertonea, D. Hooper, , and J. Silk, Phys. Rep.Volume 405,
Issues 5-6, January 2005, Pages 279-39

\bibitem{dark2} Gia Dvali, Gregory Gabadadze, Massimo Porrati,
Phys.Lett.B485:208-214,2000.

\bibitem{dark3} NS-branes in 5d brane world models, Eun Kyung Park, Pyung
Seong Kwon, arXiv:1007.1290.

\bibitem{cooper} Supersymmetry in Quantum Mechanics, F.Cooper, A. Khare, U.
Sukhatme; World Scientific Publishing (2001), ISBN 981-02-4605-6.

\bibitem{monopolospinsusyA} Mikhael S. Plyushchay, Phys.Lett. B485 (2000)
187; Sergey M. Klishevich and Mikhail S. Plyushchay, Nucl. Phys. B616 (2001)
403, and Conference on Symmetry in Nonlinear Mathematical Physics, Kiev,
Ukraine, 9-15 Jul 2001.

\bibitem{monopolosusyC} Eric D'Hoker, Lue Vinet, Phys. Lett. 137 B 1 2
(1984) 72.

\bibitem{ideiainicial} F. de Jonghe, A. J. Macfarlane, K. Peeters, J.W.Van
Holten, Phys. Lett. B 359 (1995) 114.

\bibitem{spinemsusy} Michael Stone, Nucl. Phys. B 314 (1989) 557.

\bibitem{dimo} S.Dimopoulos, D.Eichler, R.Esmailzadeh and G.D.Starkman,
Phys.Rev. D \ 41, 2388 (1990).)
\end{thebibliography}
\end{document}